\documentclass[showpacs, amsmath,amssymb, aps,showkeys]{revtex4}
\usepackage{amsmath,amssymb,amsfonts,latexsym,float,graphics,epsfig,bm}
\usepackage{graphicx}

\catcode`@=11 \@addtoreset{equation}{section}

\newcommand{\be}{\begin{equation}}
\newcommand{\ee}{\end{equation}}

\begin{document}
\title{The Tomimatsu-Sato metric reloaded}
\author{D. Batic}
\email{davide.batic@ku.ac.ae}
\affiliation{
Department of Mathematics,\\  Khalifa University of Science and Technology,\\ Main Campus, Abu Dhabi,\\ 
United Arab Emirates}

\date{\today}
\begin{abstract}
In this work, we derive exact analytic formulae for the inner and outer surfaces representing the boundary of the ergoregion appearing in the Tomimatsu-Sato (TS) metric. Exact expressions for the radii of the ergoregion in prolate spheroidal coordinates and in Boyer-Lindquist coordinates are obtained. We also found that in addition to the ring-shaped naked singularity there is an event horizon placed in the inner region inside the aforementioned curvature singularity. In comparing our results with previous studies, we also uncovered
and corrected several errors in the literature. Finally, we provide tables of numerical values for the inner and outer boundaries of the ergoregion for different values of the rotational parameter. We hope this study will be a useful resource for all researchers interested in the Tomimatsu-Sato metric.
\end{abstract}


\maketitle

\section{Introduction}
In 1972, nine years after Kerr obtained the metric associated to a spinning mass, Tomimatsu and Sato were able to derive a new solution to the Einstein field equations representing a stationary axisymmetric and asymptotically flat metric describing the geometry around a deformed spinning mass with deformation parameter $\delta=2$ \cite{tomsato,tomsatop}. Such a solution was gradually generalized to the case of an arbitrary positive integer distortion parameter in the series of papers \cite{tomsatop,Yam01,Yam02,Hori,Tani}. In the present work, we are interested in the case $\delta=2$ for which \cite{gary} showed that the corresponding space-time is characterized by four distinct principal null directions and therefore, it is of type I according to Petrov's classification. We recall that the Kerr metric is instead of type D and as such it admits a Carter constant which plays a fundamental role in the separation process and reduction of order for the geodesic equation while the TS metric being of general Petrov type will not allow for Carter's integral of motion \cite{CH}. Another striking difference between the aforementioned manifolds is that differently as in the Kerr metric where an event horizon shields the ring-like curvature singularity, the TS space-time has a naked spinning singularity \cite{gary} in vicinity of which closed time-like curves appear \cite{gary}. If from one side this feature is at odd with the Cosmic Censorship Conjecture (CCC) \cite{Penrose} and we may be induced to dismiss the TS solution as unphysical, on the other side, we should be aware that CCC is still a conjecture and up to now, several counterexamples are known \cite{Papa1,Hollier,Singh,Gundlach,Harada,Boyda,Israel,Drukker}.

There has been a long debate in the literature on whether the TS metric may admits an event horizon. 
Both Kerr and TS have an ergoregion but TS has a ring-like curvature singularity at the intersection of the inner surface of the ergoregion with the equatorial plane.
Everybody uses prolate spheroidal coordinates while we use Boyer-Lindquist coordinates adapted to the present problem since Kerr metric also usually written in terms of BL-like coordinates and this makes easier to higlight similarities and differences between these two metrics.

One of the reasons leading us to study the TS metric is that the literature on this topic is characterized by several mistakes, misprints happening from the derivation of the TS metric and extending until Bambi where the shadow of a TS gravitational object with $\delta=2$ is studied and instead of using the BL coordinates corresponding to that case which can be found in Yamazaki Bambi took the BL coordinates for $\delta=1$, i.e. the Kerr metric. This led us to go over the derivation of the TS metric in Section 2 where we pointed out some mistakes and typos occurring in the literature. In section 3, we focused our attention on the study of null surfaces. We found that in addition to the ring-shaped naked singularity there is an event horizon placed in the inner region inside the aforementioned curvature singularity. In Section 4, we draw our conclusions and  discuss future research directions related to the TS metric and its possible extension in the presence of a positive cosmological constant.

\section{Derivation of the Tomimatsu-Sato metric with deformation parameter $\delta=2$}
The general form of the Lewis-Papapetrou line element describing cylindrical solutions to the vacuum Einstein field equations in cylindrical coordinates $(x^0,x^1,x^2,x^3)=(t,\rho,z,\varphi)$ is \cite{Lewis,Papa}
\begin{equation}\label{Lewis-Papa}
ds^2=f dt^2-2\kappa dtd\varphi-\ell d\varphi^2-e^\mu(d\rho^2+dz^2),
\end{equation}
where the four unknown functions appearing in (\ref{Lewis-Papa}) depend on the variables $\rho$ and $z$ only. In the reminder of this section, we will assume that the metric coefficients satisfy the condition $\partial_{\beta\alpha}g_{\mu\nu}=\partial_{\alpha\beta}g_{\mu\nu}$. Note that if we want the line element (\ref{Lewis-Papa}) to go over to the Minkowski metric in cylindrical coordinates as $\rho,z\to\infty$, i.e.
\begin{equation}
ds^2=dt^2-\rho^2 d\varphi^2-d\rho^2-dz^2,
\end{equation}
we need to require in such limit that $f\to 1$, $\kappa,\mu\to 0$ and $\ell\to\rho^2$. If we impose that the Ricci tensor $R_{\alpha\beta}$ vanishes, we find by means of Maple that the only components that do not identically vanish are $R_{00}$, $R_{03}$, $R_{11}$, $R_{12}$, $R_{22}$ and $R_{33}$. More precisely, we end up with the following overdetermined coupled system of PDEs
\begin{eqnarray}
\partial_\rho\left(\frac{\partial_\rho f}{D}\right)+\partial_z\left(\frac{\partial_z f}{D}\right)+\frac{f}{D^3}\left[\partial_\rho f\partial_\rho\ell+\partial_z f\partial_z\ell+(\partial_\rho\kappa)^2+(\partial_z\kappa)^2\right]&=&0,\label{2.2a}\\
\partial_\rho\left(\frac{\partial_\rho\kappa}{D}\right)+\partial_z\left(\frac{\partial_z \kappa}{D}\right)+\frac{\kappa}{D^3}\left[\partial_\rho f\partial_\rho\ell+\partial_z f\partial_z\ell+(\partial_\rho\kappa)^2+(\partial_z\kappa)^2\right]&=&0,\label{2.2b}\\
\partial_\rho\left(\frac{\partial_\rho\ell}{D}\right)+\partial_z\left(\frac{\partial_z \ell}{D}\right)+\frac{\ell}{D^3}\left[\partial_\rho f\partial_\rho\ell+\partial_z f\partial_z\ell+(\partial_\rho\kappa)^2+(\partial_z\kappa)^2\right]&=&0,\label{2.2c}\\
\partial_{\rho\rho}\mu+\partial_{zz}\mu+2\frac{\partial_{\rho\rho}D}{D}-\frac{1}{D}\left(\partial_\rho\mu\partial_\rho D-\partial_z\mu\partial_z D\right)-\frac{1}{D^2}\left[\partial_\rho f\partial_\rho\ell+\left(\partial_\rho\kappa\right)^2\right]&=&0,\label{R11}\\
\partial_{\rho\rho}\mu+\partial_{zz}\mu+2\frac{\partial_{zz}D}{D}+\frac{1}{D}\left(\partial_\rho\mu\partial_\rho D-\partial_z\mu\partial_z D\right)-\frac{1}{D^2}\left[\partial_z f\partial_z\ell+\left(\partial_z\kappa\right)^2\right]&=&0,\label{R22}\\
\partial_z\left(\frac{\partial_\rho D}{D}\right)-\frac{1}{2D}\left(\partial_z\mu\partial_\rho D+\partial_\rho\mu\partial_z D\right)+\frac{\Psi}{4D^4}&=&0,\label{R12}
\end{eqnarray}
with 
\begin{eqnarray}
D^2&=&\kappa^2+f\ell,\\
\Psi&=&\kappa^2\left(2\partial_\rho\kappa\partial_z\kappa-\partial_\rho f\partial_z\ell-\partial_z f\partial_\rho\ell\right)+2\kappa\ell\left(\partial_z\kappa\partial_\rho f+\partial_\rho\kappa\partial_z f\right)+2\kappa f\left(\partial_\rho\kappa\partial_z\ell+\partial_z\kappa\partial_\rho\ell\right)\nonumber\\
&-&2f\ell\partial_\rho\kappa\partial_z\kappa+f^2\partial_\rho\ell\partial_z\ell+\ell^2\partial_\rho f\partial_z f.
\end{eqnarray}
Note that equations (\ref{2.2a}), (\ref{2.2b}) and (\ref{2.2c}) agree with the corresponding equations (2.2a), (2.2b) and (2.2c) in \cite{Islam} where the corresponding equations arising from the Ricci components $R_{11}$, $R_{12}$ and $R_{22}$ have not been given. As already observed by \cite{Lewis}, the combination $\ell R_{00}-2\kappa R_{03}-fR_{33}$ gives rise to the Laplace equation $\partial_{\rho\rho}D+\partial_{zz}D=0$. This is a remarkable property satisfied by the function $D$ which leads to a further simplification of the line element (\ref{Lewis-Papa}). More precisely, $D$ is the real part of some complex-valued function $\Sigma(\rho+iz)=D(\rho,z)+iE(\rho,z)$ where the functions $D$ and $E$ are intertwined through the Cauchy-Riemann equations $\partial_\rho D=\partial_z E$ and $\partial_z D=-\partial_\rho E$. If we introduce the coordinate transformation 
\begin{equation}\label{2.6}
\overline{\rho}=D(\rho,z),\quad\overline{z}=E(\rho,z)
\end{equation}
together with the invertibility condition $\partial_\rho D\partial_z E-\partial_\rho E\partial_z D=(\partial_\rho D)^2+(\partial_z D)^2\neq 0$, we can rewrite the line element (\ref{Lewis-Papa}) as
\begin{equation}\label{newm}
ds^2=f dt^2-2\kappa dtd\varphi-\ell d\varphi^2-e^{\overline{\mu}}(d\overline{\rho}^2+d\overline{z}^2),\quad
e^{\overline{\mu}}=\frac{e^\mu}{(\partial_\rho D)^2+(\partial_z D)^2},
\end{equation}
where the functions $f$, $\kappa$ and $\ell$ depend on the new variables $\overline{\rho}$ and $\overline{z}$. At this point, we can exploit the freedom of choosing $D$ so that it satisfies the Laplace equation. The simplest choice is $D=\rho$ which implies that $\overline{\rho}=\rho$ while the Cauchy-Riemann equations lead to $E=z$ modulo an integration constant that can be chosen to be zero. Hence, we have $\overline{z}=z$ and as a consequence, the transformed metric (\ref{newm}) coincides again with (\ref{Lewis-Papa}) but the unknown functions $f$, $\kappa$ and $\ell$ are not anymore independent because they are connected through the relation
\begin{equation}\label{relD}
D^2=\kappa^2+f\ell=\rho^2.
\end{equation}
If we implement (\ref{relD}) into (\ref{2.2a})-(\ref{R12}), the resulting Einstein field equations read
\begin{eqnarray}
\partial_{\rho\rho}f+\partial_{zz}f-\frac{\partial_\rho f}{\rho}+\frac{f}{\rho^2}\left[\partial_\rho f\partial_\rho\ell+\partial_z f\partial_z\ell+(\partial_\rho\kappa)^2+(\partial_z\kappa)^2\right]&=&0,\label{*}\\
\partial_{\rho\rho}\kappa+\partial_{zz}\kappa-\frac{\partial_\rho\kappa}{\rho}+\frac{\kappa}{\rho^2}\left[\partial_\rho f\partial_\rho\ell+\partial_z f\partial_z\ell+(\partial_\rho\kappa)^2+(\partial_z\kappa)^2\right]&=&0,\label{**}\\
\partial_{\rho\rho}\ell+\partial_{zz}\ell-\frac{\partial_\rho\ell}{\rho}+\frac{\ell}{\rho^2}\left[\partial_\rho f\partial_\rho\ell+\partial_z f\partial_z\ell+(\partial_\rho\kappa)^2+(\partial_z\kappa)^2\right]&=&0,\label{***}\\
\partial_{\rho\rho}\mu+\partial_{zz}\mu-\frac{\partial_\rho\mu}{\rho}-\frac{1}{\rho^2}\left[\partial_\rho f\partial_\rho\ell+\left(\partial_\rho\kappa\right)^2\right]&=&0,\label{2.10a}\\
\partial_z\mu+\frac{1}{2\rho}\left(\partial_\rho f\partial_z\ell+\partial_z f\partial_\rho\ell+2\partial_\rho\kappa\partial_z\kappa\right)&=&0,\label{2.10b}\\
\partial_{\rho\rho}\mu+\partial_{zz}\mu+\frac{\partial_\rho\mu}{\rho}-\frac{1}{\rho^2}\left[\partial_z f\partial_z\ell+\left(\partial_z\kappa\right)^2\right]&=&0.\label{2.10c}
\end{eqnarray}
As a consistency check, it is gratifying to observe that equations (\ref{*})-(\ref{***}) coincide with equations (A)-(C) in \cite{Papa}. Moreover, (\ref{2.10a})-(\ref{2.10c}) agree with (2.10a)-(2.10b) in \cite{Islam}, respectively. Finally, if we consider the combinations (\ref{2.10a})$\pm$(\ref{2.10c}), we obtain equations (2.2) and (2.2a) in \cite{Papa}, namely
\begin{eqnarray}
\partial_{\rho\rho}\mu+\partial_{zz}\mu&=&\frac{1}{2\rho^2}\left[\partial_\rho f\partial_\rho\ell+\partial_z f\partial_z\ell+(\partial_\rho\kappa)^2+(\partial_z\kappa)^2\right],\\
\partial_\rho\mu&=&-\frac{1}{2\rho}\left[\partial_\rho f\partial_\rho\ell-\partial_z f\partial_z\ell+(\partial_\rho\kappa)^2-(\partial_z\kappa)^2\right].\label{*****}
\end{eqnarray}
Note that equations (\ref{*})-(\ref{***}) control the functions $k$, $f$ and $\ell$ and once they are found, we can compute $\mu$ by quadratures from (\ref{2.10b}) and (\ref{*****}). In order to derive the Tomimatsu-Sato metric with deformation parameter $\delta=2$, it is convenient to recast (\ref{Lewis-Papa}) in the so-called Weyl-Lewis-Papapetrou form. This is achieved with the help of (\ref{relD}) and by introducing the new function
\begin{equation}
w=\frac{\kappa}{f}.
\end{equation}
As a result, the line element (\ref{Lewis-Papa}) becomes
\begin{equation}\label{2.16}
ds^2=f(dt-wd\varphi)^2-\frac{\rho^2}{f}d\varphi^2-e^\mu(d\rho^2+dz^2)
\end{equation}
The corresponding Einstein field equations can be readily obtained from (\ref{*}), (\ref{**}), (\ref{2.10b}) and (\ref{*****}). They are
\begin{eqnarray}
&&f\left(\partial_{\rho\rho}f+\partial_{zz}f+\frac{\partial_\rho f}{\rho}\right)-\left(\partial_\rho f\right)^2-\left(\partial_z f\right)^2+\frac{f^4}{\rho^2}\left[\left(\partial_\rho w\right)^2+\left(\partial_z w\right)^2\right]=0,\label{2.12a}\\
&&f\left(\partial_{\rho\rho}w+\partial_{zz}w-\frac{\partial_\rho w}{\rho}\right)+2\left(\partial_\rho w\partial_\rho f+\partial_z w\partial_z f\right)=0,\label{2.12b}\\
&&\partial_\rho\mu=-\frac{\partial_\rho f}{f}+\frac{\rho}{2f^2}\left[\left(\partial_\rho f\right)^2-\left(\partial_z f\right)^2\right]-\frac{f^2}{2\rho}\left[\left(\partial_\rho w\right)^2-\left(\partial_z w\right)^2\right],\label{2.13a}\\
&&\partial_z\mu=-\frac{\partial_z f}{f}+\frac{\rho}{f^2}\partial_\rho f\partial_z f-\frac{f^2}{\rho}\partial_\rho w\partial_z w.\label{2.13b}
\end{eqnarray}
At this point a couple of remarks are necessary. First of all, it is not difficult to check that the equations (\ref{2.13a}) and (\ref{2.13b}) are consistent, i.e. $\partial_{\rho z}\mu=\partial_{z\rho}\mu$. Moreover, $\mu$ defined through (\ref{2.13a}) and (\ref{2.13b}) satisfies the equations (\ref{2.10a}) and (\ref{2.10c}). This can be verified by replacing $\kappa=wf$ and $\ell=\rho^2/f-fw^2$ therein. Finally, equation (\ref{***}) after the aforementioned substitution coincides with (\ref{2.12a}). Hence, the new Einstein field equations are represented by (\ref{2.12a})-(\ref{2.13b}). It is worth mentioning that \cite{Papa} solved the above system of equations under the additional assumption that the r.h.s. of (\ref{2.12b}) vanishes. Such an approach leads to a solution describing a gravitational field where either the mass or the angular momentum can be different from zero. Last but not least, instead of working with the functions $f$ and $w$,  \cite{Cosgrove1,Cosgrove2} introduced a new function $\gamma=(\mu+\log{f})/2$ and derived a fourth order quasi-linear PDE for $\gamma$. A less restrictive procedure than the one adopted in \cite{Papa} relies on the observation that the divergence of the vector field 
$\mathcal{F}:\Omega\subseteq (0,\infty)\times\mathbb{R}\longrightarrow\mathbb{R}^2$ defined as follows
\begin{equation}
\mathcal{F}(\rho,z)=\frac{f^2}{\rho}\left(\partial_\rho w,\partial_z w\right)
\end{equation}
coincides with the LHS of (\ref{2.12b}). This is easily done by showing that the equation $\partial_\rho(\mathcal{A}\partial_\rho w)+\partial_z(\mathcal{A}\partial_z w)=0$ coincides with (\ref{2.12b}) if $\mathcal{A}=f^2/\rho$. Hence, it is possible to construct a function $u=u(\rho,z)$ such that
\begin{equation}\label{3.2}
\partial_\rho u=\frac{f^2}{\rho}\partial_z w,\quad
\partial_z u=-\frac{f^2}{\rho}\partial_\rho w.
\end{equation}
By means of (\ref{3.2}) it is not difficult to derive from (\ref{2.12a})-(\ref{2.13b}) the following PDEs governing the functions $f$ and $u$, namely
\begin{eqnarray}
f\nabla^2 f&=&\left(\partial_\rho f\right)^2+\left(\partial_z f\right)^2-\left[\left(\partial_\rho u\right)^2+\left(\partial_z u\right)^2\right],\label{3.3a}\\
f\nabla^2 u&=&2\left(\partial_\rho f\partial_\rho u+\partial_z f\partial_z u\right),\label{3.3b}\\
\partial_\rho\left(\mu+\ln{f}\right)&=&\frac{\rho}{2f^2}\left[\left(\partial_\rho f\right)^2-\left(\partial_z f\right)^2\right]+\frac{\rho}{2f^2}\left[\left(\partial_\rho u\right)^2-\left(\partial_z u\right)^2\right],\label{3.4a}\\
\partial_z\left(\mu+\ln{f}\right)&=&\frac{\rho}{f^2}\left(\partial_\rho f\partial_z f+\partial_\rho u\partial_z u\right),\label{3.4b}
\end{eqnarray}
where $\nabla^2=\rho^{-1}\partial_\rho(\rho\partial_\rho\cdot))+\rho^{-2}\partial_{\varphi\varphi}+\partial_{zz}$ is the Laplace operator in cylindrical coordinates. We draw the attention of the reader to the fact that there is a typo in equation (3.3a) in \cite{Islam} where the term $\partial_z u$ should appear squared. The Ernst equation \cite{Ernst} emerges from (\ref{3.3a}) and (\ref{3.3b}) by introducing some complex-valued function $\mathcal{E}=f+iu$ and realizing that the aforementioned equations are the real and imaginary parts of the complex PDE
\begin{equation}\label{Ernst}
\Re{\left(\mathcal{E}\right)}\nabla^2\mathcal{E}=\left(\partial_\rho\mathcal{E}\right)^2+\left(\partial_z\mathcal{E}\right)^2,\quad\Re{\mathcal{E}}=f.
\end{equation}
Hence, (\ref{Ernst}) is an equivalent form of the Einstein field equations (\ref{3.3a}) and (\ref{3.3b}). In order to derive the Tomimatsu-Sato metric, it is convenient to introduce the Ansatz
\begin{equation}\label{3.7}
\mathcal{E}=\frac{\Phi-1}{\Phi+1}
\end{equation}
with $\Phi$ a complex-valued function yet to be determined. The Ernst equation becomes
\begin{equation}\label{3.8}
\left(|\Phi|^2-1\right)\nabla^2\Phi=2\Phi^{*}\left[\left(\partial_\rho\Phi\right)^2+\left(\partial_z\Phi\right)^2\right],
\end{equation}
where star denotes complex conjugation. Note that our notation departs from that in \cite{Islam,tomsato,tomsatop} where the lowercase Greek  letter $\xi$ is used instead of $\Phi$. The reason behind our choice is that the letter $\xi$ will denote one of the prolate spheroidal coordinates. Finally, the Tomimatsu-Sato line element can be constructed by searching for solutions to (\ref{3.8}) in prolate spheroidal coordinates $\xi=\xi(\rho,z)$, $\eta=\eta(\rho,z)$ with $\xi\geq 1$ and $-1\leq \eta\leq 1$ such that
\begin{equation}\label{3.9}
\rho=\sigma\sqrt{(\xi^2-1)(1-\eta^2)},\quad
z=\sigma\xi\eta,\quad\sigma>0.
\end{equation}
The corresponding inversion formulae are
\begin{equation}\label{3.10}
\xi=\frac{\sqrt{R_+}+\sqrt{R_{-}}}{2\sigma},\quad\eta=\frac{\sqrt{R_+}-\sqrt{R_{-}}}{2\sigma},\quad
R_\pm=\rho^2+(z\pm\sigma)^2.
\end{equation}
We alert the reader that our notation for the prolate spheroidal coordinates $(\xi,\eta)$ differs from that employed in \cite{Islam,tomsato,tomsatop} where the lowercase Latin letters $x$ and $y$ stand for $\xi$ and $\eta$, respectively. If we apply the coordinate transformation (\ref{3.9}) to (\ref{3.8}), we end up with the Ernst equation in prolate spheroidal coordinates, i.e.
\[
\left(|\Phi|^2-1\right)\left[(\xi^2-1)\partial_{\xi\xi}\Phi+(1-\eta^2)\partial_{\eta\eta}\Phi+2\xi\partial_\xi\Phi-2\eta\partial_\eta\Phi\right]=
\]
\begin{equation}\label{3.11}
2\Phi^{*}\left[(\xi^2-1)\left(\partial_\xi\Phi\right)^2+(1-\eta^2)\left(\partial_\eta\Phi\right)^2\right].
\end{equation}
Moreover, the equations (\ref{3.2}), (\ref{3.4a}) and (\ref{3.4b}) become
\begin{eqnarray}
\partial_\xi  w&=&-\frac{\sigma(1-\eta^2)}{f^2}\partial_\eta u,\label{3.12I}\\
\partial_\eta w&=&\frac{\sigma(\xi^2-1)}{f^2}\partial_\xi u,\label{3.12II},\\
\partial_\xi\left(\mu+\ln{f}\right)&=&\frac{1-\eta^2}{2f^2(\xi^2-\eta^2)}
\left\{\xi(\xi^2-1)\left[\left(\partial_\xi f\right)^2+\left(\partial_\xi u\right)^2\right]-\xi(1-\eta^2)\left[\left(\partial_\eta f\right)^2+\left(\partial_\eta u\right)^2\right]\right.\nonumber\\
&-&\left.2\eta(\xi^2-1)\left(\partial_\xi f\partial_\eta f+\partial_\xi u\partial_\eta u\right)\right\},\label{3.13a}\\
\partial_\eta\left(\mu+\ln{f}\right)&=&\frac{\xi^2-1}{2f^2(\xi^2-\eta^2)}
\left\{\eta(\xi^2-1)\left[\left(\partial_\xi f\right)^2+\left(\partial_\xi u\right)^2\right]-\eta(1-\eta^2)\left[\left(\partial_\eta f\right)^2+\left(\partial_\eta u\right)^2\right]\right.\nonumber\\
&+&\left.2\xi(1-\eta^2)\left(\partial_\xi f\partial_\eta f+\partial_\xi u\partial_\eta u\right)\right\}.\label{3.13b}
\end{eqnarray}
It is well-known that the vacuum stationary axisymmetric solutions to Einstein's field equations can be generated from the Ernst equation (\ref{Ernst}), (\ref{3.8}) or equivalently from (\ref{3.11}). For instance, the Kerr metric emerges from the case
\begin{equation}
\Phi_K=p\xi-iq\eta,\quad q=\frac{J}{M^2},\quad p=\sqrt{1-q^2},
\end{equation}
where $M$ is the mass of the gravitational source and $J$ its total angular momentum. A further solution to (\ref{3.11}) was found by \cite{tomsato,tomsatop} in the form
\begin{equation}\label{phits}
\Phi_{TS}=\frac{\mathfrak{u}+i\mathfrak{v}}{\mathfrak{m}+i\mathfrak{n}}
\end{equation}
with
\begin{equation}
\mathfrak{u}=p^2 \xi^4+q^2\eta^4-1,\quad
\mathfrak{v}=-2pq\xi\eta(\xi^2-\eta^2),\quad
\mathfrak{m}=2p\xi(\xi^2-1),\quad
\mathfrak{n}=-2q\eta(1-\eta^2).
\end{equation}
It is a straightforward exercise with Maple to verify that (\ref{phits}) indeed satisfies (\ref{3.11}). Since $\Phi_{TS}\to (\xi^2+1)/2\xi$ for $q\to 0$, in this limit the Tomimatsu-Sato solution coincides with the class of Weyl's metrics generated by 
\begin{equation}
\Phi_W=\frac{(\xi+1)^\delta+(\xi-1)^\delta}{(\xi+1)^\delta-(\xi-1)^\delta}
\end{equation}
when $\delta=2$. In the case $\delta=1$, the function $\Phi_W$ leads to the Schwarzschild metric \cite{Voor}. Therefore,  $\delta$ and $q$ can be viewed as positive parameters measuring the deviation from spherical symmetry. If we insert (\ref{phits}) into (\ref{3.7}) and recall that $\mathcal{E}=f+iu$, we can easily compute the real and imaginary parts of $\mathcal{E}$ with the help of Maple. More precisely, we find that
\begin{equation}
f=\Re{\left(\mathcal{E}\right)}=\frac{A}{B}~,\quad u=\Im{\left(\mathcal{E}\right)}=\frac{2I}{B}
\end{equation}
with
\begin{eqnarray}
A&=&\mathfrak{u}^2+\mathfrak{v}^2-(\mathfrak{m}^2+\mathfrak{n}^2),\\
&=&\left[p^2(\xi^2-1)^2+q^2(1-\eta^2)^2\right]^2-4p^2 q^2(\xi^2-1)(1-\eta^2)(\xi^2-\eta^2)^2,\label{Acof}\\
B&=&(\mathfrak{u}+\mathfrak{m})^2+(\mathfrak{v}+\mathfrak{n})^2,\\
&=&(p^2\xi^4+q^2\eta^4-1+2p\xi^3-2p\xi)^2+4q^2\eta^2(p\xi^3-p\xi\eta^2+1-\eta^2)^2,\label{Bcof}\\
I&=&\mathfrak{m}\mathfrak{v}-\mathfrak{n}\mathfrak{u},\\
&=&-2q\eta\left[(1-\eta^2)(1-q^2\eta^4)+p^2\xi^2\eta^2(2-\xi^2)+p^2\xi^4(2\xi^2-3)\right].
\end{eqnarray}
It is gratifying to observe that the expressions for the coefficients $A$ and $B$ represented by (\ref{Acof}) and (\ref{Bcof}) agree with the corresponding ones offered in\cite{Islam,tomsato,tomsatop}. Moreover, $f$ exhibits the desired behaviour at space-like infinity, that is $f\to 1$ as $\xi\to\infty$. Concerning the function $w$, it can derived from the equations (\ref{3.12I}) and (\ref{3.12II}). In particular, we try as in \cite{tomsato,tomsatop,Islam} the Ansatz
\begin{equation}\label{answ}
w=\frac{2Mq(1-\eta^2)}{A}C,\quad C=\sum_{\substack{k,l=0\\ 0\leq k+l\leq 7}}^7 a_{kl}\xi^k\eta^l,
\end{equation}
where $M$ is the total mass of the gravitational source and  $C$ is some polynomial in $\xi$ and $\eta$ of maximum degree seven. This requirement together with the fact that $A$ is a polynomial of degree eight in $\xi$ is necessary in order to ensure that $w\to 0$ as $\xi\to\infty$. If we substitute (\ref{answ}) into (\ref{3.12I}) and (\ref{3.12II}), we obtain the following system
\begin{eqnarray}
A\partial_\xi C-C\partial_\xi A&=&\frac{\sigma}{Mq}\left(I\partial_\eta B-B\partial_\eta I\right),\label{Cpde1}\\
(1-\eta^2)(A\partial_\eta C-C\partial_\eta A)-2\eta AC&=&\frac{\sigma}{Mq}(\xi^2-1)\left(B\partial_\xi I-I\partial_\xi B\right)\label{Cpde2}
\end{eqnarray}
in agreement with equations (7) and (8) in \cite{Yam01}. Setting $\sigma=Mp/2$ and substituting (\ref{answ}) into (\ref{Cpde1}) leads to a linear system for the unknown coefficients $a_{kl}$ which is easily handled by Maple. The final result is
\[
C=q^2(1+p\xi)(1-\eta^2)^3-p^2(\xi^2-1)(1-\eta^2)(p\xi^3+3\xi^2+3p\xi+1)
\]
\begin{equation}\label{Cof}
-2p^2\xi(\xi^2-1)^2(p\xi^2+2\xi+p).
\end{equation}
As a consistency check, we used Maple to verify that (\ref{Cof}) is indeed a solution to both PDEs (\ref{Cpde1}) and (\ref{Cpde2}). It is gratifying to see that (\ref{Cof}) agrees with the corresponding expressions given in \cite{tomsatop,Yam01}. However, a comment is in order. \cite{tomsatop} also derived a system of first order PDEs for $C$ represented by equations (3.3) and (3.4) therein. Even though the l.h.s.'s of (\ref{Cpde1}) and (3.3) in \cite{tomsatop} coincide, the same cannot be said for the r.h.s.'s. Despite the fact that the expression for $C$ given in \cite{tomsatop} agrees with the corresponding expressions presented here and in \cite{Yam01}, it does not satisfy equation (3.3) in \cite{tomsatop} as it can be easily verified by means of Maple. The only conclusion is that such a discrepancy is due to a typo in the aforementioned equation. Finally, \cite{Islam} instead of deriving a formula for $C$ gives the same result as in \cite{tomsatop} but with a typo in the last term. It should also be mentioned that \cite{Islam} makes an opposite choice for the signs entering in front of the l.h.s. of the equations represented by (\ref{3.2}) in the present work. It is not difficult to check that such a choice would alter the sign in front of the l.h.s.'s in (\ref{Cpde1}) and (\ref{Cpde2}) and send $C$ into $-C$. A formula for the metric coefficient $e^{2\gamma}/f$ was derived in \cite{tomsatop} while \cite{Islam} gives an expression for $e^\mu$ which is linked to the aforementioned metric coefficient in \cite{tomsatop} by the relation $e^\mu=e^{2\gamma}/f$. We checked with Maple that 
\begin{equation}\label{mu}
e^\mu=\frac{B}{p^4(\xi^2-\eta^2)^4}
\end{equation}
is indeed a solution of the system (\ref{3.13a}) and (\ref{3.13b}). Finally, the line element (\ref{2.16}) can be written in prolate spheoridal coordinates as follows
\begin{eqnarray}
ds^2=\frac{A}{B}dt^2&-&\frac{4Mq(1-\eta^2)C}{B}dtd\varphi-\frac{M^2 B}{4p^2(\xi^2-\eta^2)^3}\left(\frac{d\xi^2}{\xi^2-1}+\frac{d\eta^2}{1-\eta^2}\right)\nonumber\\
&-&\frac{M^2(1-\eta^2)}{A}\left[\frac{p^2}{4}(\xi^2-1)B-4q^2(1-\eta^2)\frac{C^2}{B}\right]d\varphi^2,\label{LEpsc}
\end{eqnarray}
where $A$, $B$ and $C$ are given by (\ref{Acof}), (\ref{Bcof}) and (\ref{Cof}), respectively. If, instead, we use the standard  definition of prolate spheoridal coordinates $(u,v,\varphi)$ with $\xi=\cosh{u}$ and $\eta=\cos{v}$ where $u\geq 0$ and $0\leq v\leq\pi$, (\ref{LEpsc}) can be cast into the form 
\[
ds^2=\frac{A}{B}dt^2-4Mq\sin^2{v}\frac{C}{B}dtd\varphi-\frac{M^2 B}{4p^2(\cosh^2{u}-\cos^2{v})^3}\left(du^2+dv^2\right)
\]
\begin{equation}\label{LEpsc1}
-\frac{M^2\sin^2{v}}{4AB}\left(p^2 B^2\sinh^2{u}-16q^2C^2\sin^2{v}\right)d\varphi^2,
\end{equation}
where
\begin{eqnarray}
A&=&\left(p^2\sinh^4{u}+q^2\sin^4{v}\right)^2-4p^2 q^2\sinh^2{u}\sin^2{v}\left(\cosh^2{u}-\cos^2{v}\right)^2,\label{AAcof}\\
B&=&\left(q^2\cos^4{v}+p^2\cosh^4{u}+2p\cosh{u}\sinh^2{u}-1\right)^2\\
&+&4q^2\cos^2{v}\left(p\cosh^3{u}-p\cosh{u}\cos^2{v}+\sin^2{v}\right)^2,\label{BBcof}\\
C&=&q^2\sin^6{v}(1+p\cosh{u})-p^2\sinh^2{u}\sin^2{v}(p\cosh^3{u}+3\cosh^2{u}+3p\cosh{u}+1)\nonumber\\
&-&2p^2\cosh{u}\sinh^4{u}\left(p\cosh^2{u}+2\cosh{u}+p\right).\label{CCof}
\end{eqnarray}

\section{Analysis of the metric}
In this section, we improve the singularity and ergoregion analysis provided in \cite{gary,Kodama}. First of all, a direct inspection of (\ref{Acof}), (\ref{Bcof}), (\ref{Cof}) and (\ref{mu}) shows that all metric coefficients are invariant under the transformation $\eta\to-\eta$. The fact that $B$ is always nonnegative implies that the sign of the metric coefficient $f$ is uniquely controlled by the sign of the polynomial function $A$. Moreover, it can be immediately seen that the  functions (\ref{Acof}), (\ref{Bcof}) and (\ref{Cof}) have common roots at $\xi=1$ and $\eta=\pm 1$. If we Taylor expand $B$ around $\xi=\pm 1$ and $\eta=\pm 1$ as follows
\begin{eqnarray}
B&=&p^2(\xi^2-1)^2a_1(\xi)-4p^2q^2(\xi^2-1)^3(1-\eta^2)+a_2(\xi)(1-\eta^2)^2\\
&+&a_3(\xi)(\eta^2-1)^3+q^4(\eta^2-1)^4,\\
\end{eqnarray}
with
\begin{eqnarray}
a_1(\xi)&=&p^2\xi^4+4p\xi^3+2(1+q^2)\xi^2+4p\xi+p^2,\\ 
a_2(\xi)&=&-6p^2q^2\xi^4-4pq^2\xi^3+12p^2q^2\xi^2+12pq^2\xi+2q^2(1+3q^2),\\
a_3(\xi)&=&4p^2 q^2\xi^2+8pq^2\xi+4q^2(1+q^2)
\end{eqnarray}
and we recall that $f=A/B$, we conclude that the metric coefficient $g_{tt}$ has a regular behaviour on the line segment described by $\xi=1$ and $-1<\eta< 1$. The points $(\xi,\eta)=(1,\pm 1)$ making $f$ singular have the interpretation of quasi-regular singularities \cite{Kodama}. At this step, it is not obvious whether the polynomial functions $A$ and $B$ have some other common zeroes in addition to those mentioned above. Inspired by the parametric surface representation of the ergosphere of a Kerr black hole, we introduce the ansatz \cite{gary}
\begin{equation}\label{ansatz}
\xi^2=1+\lambda^2(1-\eta^2)
\end{equation}
with $\lambda\in\mathbb{R}$. This allows to cast $A$ into the form
\begin{equation}\label{Adar}
A=(1-\eta^2)^4\mathfrak{p}_\pm,\quad
\mathfrak{p}_\pm=p^2\lambda^4+q^2\pm 2pq\lambda(\lambda^2+1).
\end{equation}
First of all, we observe that a quartic polynomial similar to $\mathfrak{p}_{-}$ has been also given in \cite{gary}. However, there is a typo there and the factor 4 multiplying the term $\lambda(\lambda+1)$ should be replaced by 2. The polynomial $\mathfrak{p}_+$ does not admit any real root for $0<q<1$ and therefore, if there are some other real roots of $A$ distinct from those already identified above, they must come from the quartic $\mathfrak{p}_{-}$. This observation is important because it signalizes that $g_{tt}$ may not be everywhere positive definite for $\xi>1$ and $-1\leq\eta\leq 1$. In other words, for decreasing values of $\xi$ the Killing vector field $\partial_t$ may already loose its property of being time-like before $\xi$ reaches the value one and an ergoregion may arise as pointed out by \cite{gary,Kodama}. Concerning the zeroes of the polynomial $\mathfrak{p}_{-}$, there are two sign changes and Descartes' rule of signs predicts that the number of positive roots is either equal two or zero. Moreover, after the transformation $\lambda\to-\lambda$ the corresponding polynomial has no sign changes and therefore, $\mathfrak{p}_{-}$ does not have negative real roots. Finally, by applying (1.163a) and (1.163b) in \cite{Bron} the two positive real roots are
\begin{equation}\label{lambdapm}
\lambda_\pm=\frac{q}{2p}+\frac{\sqrt{\sqrt[3]{4p^2 q^2}+q^2}}{2p}\pm
\sqrt{2q^2-\sqrt[3]{4p^2 q^2}+\frac{2\sqrt[3]{q^2}(2p^2+q^2)}{\sqrt{\sqrt[3]{4p^2}+q\sqrt[3]{q}}}},
\end{equation}
to which there correspond the following values of $\xi$ determined with the help of (\ref{ansatz}), i.e.
\begin{equation}\label{ergoxi}
\xi^\pm_\mathcal{E}=\sqrt{1+\lambda_\pm^2(1-\eta^2)},
\end{equation}
where the subscript $\mathcal{E}$ stands for the ergoregion 
\begin{equation}
\mathcal{E}=\{(\xi,\eta,\varphi)\in\mathbb{R}^3~|~\xi_\mathcal{E}^{-}<\xi<\xi_\mathcal{E}^{+},~-1<u<1,~0\leq\varphi<2\pi\}.
\end{equation}
Note that on the equator, i.e. $\eta=0$, the inner and outer boundaries of the ergoregion are located at
\begin{equation}\label{nunc}
\xi^\pm_{\mathcal{E},e}=\sqrt{1+\lambda_\pm^2}.
\end{equation}
In order to investigate whether or not the metric coefficient $f$ becomes singular, it is necessary to study the zeroes of $B$. If we restrict our attention to the equatorial plane as in \cite{gary}, we end up with the problem of finding the roots of the polynomial equation
\begin{equation}
B(\xi,0)=\mathfrak{P}^2=0,\quad\mathfrak{P}=p^2\xi^4+2p\xi^3-2p\xi-1.
\end{equation}
Since $\mathfrak{P}$ exhibits only one sign change, there is only one positive real root, here denoted as $\xi_B$, and it must be a zero of order two for $B$. Moreover, the polynomial obtained by means of the transformation $\xi\to-\xi$ has three sign changes meaning that there can be three or one negative real roots. As it can be seen from Table~\ref{T1}, it turns out that $\xi_B=\xi_\mathcal{E}^{-}$ in agreement with \cite{gary} where however no numerical/analytic evidence was given for such a result. Since $\xi_\mathcal{E}^{-}$ is a simple zero for $A$, while $\xi_B$ is a root of order two for $B$, we can conclude that the metric coefficient $f$ exhibits a singularity at $\xi=\xi_B$ on the equatorial plane. We checked numerically that $A$ and $B$ have no common real roots away from the equatorial plane (see Appendix~\ref{numeric} for typical values of $\xi_\mathcal{E}^\pm$ when $\eta\neq 0$). More precisely, all zeroes of $B$ are complex whenever $\eta\neq 0$. This indicates that the metric coefficient $f$ can only have a singularity at the point $(\xi,\eta)=(\xi_B,0)$ in addition to the singular points at $(1,\pm 1)$. 
\begin{table}[ht]
\caption{Typical values of the zeroes $\lambda_{\pm}$ of the polynomial equation $\mathfrak{p}_{-}=0$ in (\ref{Adar}) with the corresponding values $\xi_\mathcal{E}^\pm$ for the inner and outer boundaries of the ergoregion on the equatorial plane $\eta=0$.  The last column represents the only positive real root of order two for the function $B$ restricted to the equatorial plane.}
\begin{center}
\begin{tabular}{ | l | l | l | l|l|l|l|}
\hline
$\eta=0$ &$\lambda_{-}$ & $\xi_\mathcal{E}^{-}$&$\lambda_{+}$ & $\xi_\mathcal{E}^+$  & $\xi_B$ \\ \hline
$q=0.1$  &0.050157194   &1.001257082 &0.645657713   &1.190325117 &1.001257082\\ \hline
$q=0.2$  &0.101280894   &1.005115824 &0.877134836   &1.330174996&1.005115824\\ \hline
$q=0.3$  &0.154462485   &1.011859012 &1.085464940   &1.475884188&1.011859012\\ \hline
$q=0.4$  &0.211086895   &1.022036045 &1.301672592   &1.641448000&1.022036045\\ \hline
$q=0.5$  &0.273120641   &1.036626685 &1.547108167   &1.842157344&1.036626685\\ \hline
$q=0.6$  &0.343701531   &1.057417014 &1.849576162   &2.102601241&1.057417014\\ \hline
$q=0.7$  &0.428584499   &1.087972735 &2.260794743   &2.472082699&1.087972735\\ \hline
$q=0.8$  &0.540664408   &1.136801655 &2.909501150   &3.076556019&1.136801655\\ \hline
$q=0.9$  &0.721912637   &1.233352284 &4.299249247   &4.414016774&1.233352284\\ \hline
$q=0.99$ &1.338695474   &1.670959476 &14.08894707   &14.12439130&1.670959476\\ \hline
\end{tabular}
\label{T1}
\end{center}
\end{table}
If we recall that prolate spheroidal and cylindrical coordinates are connected by (\ref{3.9}), it can be easily checked that $(\xi_B,0)$ corresponds to a ring singularity with radius
\begin{equation}
\rho_s=\frac{Mp}{2}\lambda_{-}.
\end{equation}
This is where the inner boundary of the ergoregion intersects the equatorial plane. Hence, such a singularity resides inside the ergoregion. The ring singularity was proved to be a curvature singularity by analysing the behaviour of the Weyl curvature invariants there (see \cite{Kodama} for a detailed discussion). \cite{Kodama} also showed that such a ring singularity has zero Komar mass: an unexpected result if we think that an axisymmetric space-time such as the one described by a Kerr black hole has instead a positive Komar mass \cite{Wald} despite the fact it also exhibits a ring singularity. This aspect of the TS metric and its physical implications seem not to have been addressed in the related existing literature and they are definitely worth being studied in more detail elsewhere. Finally, the ring singularity is naked because it is not hidden by a Cauchy or event horizon as it has been already discussed in \cite{Kodama} (see Fig.~10 therein).

\begin{figure}[ht]\label{hic_02}
\includegraphics[scale=0.62]{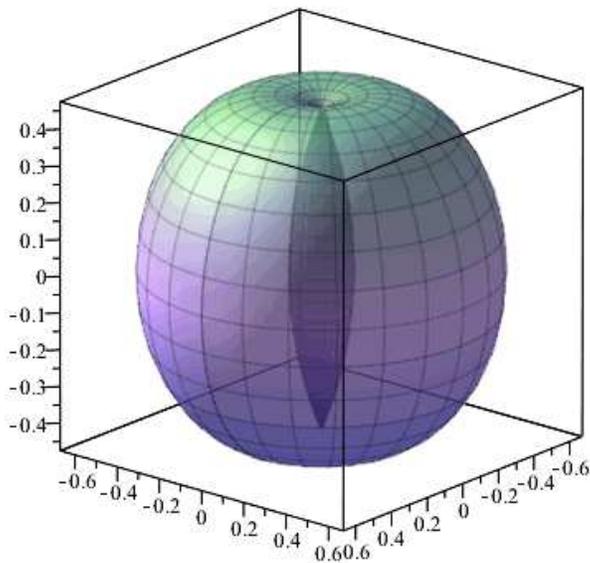}
\caption{\label{f02}
Plot of of the ergoregion for $q=0.5$ and $M=1$. Cusp singularities emerge at the top and bottom of the inner surface of the ergoregion.}
\end{figure}

It is interesting to understand how the aforementioned singularities are mapped when we switch from prolate to Boyer-Lindquist (BL) coordinates. When we reviewed the existing literature, we observed that there has been some confusion regarding the transformations from cylindrical to BL coordinates in the presence of an arbitrary deformation parameter. For instance, \cite{tomsato} gives without proof the formulae
\begin{eqnarray}
\rho&=&\frac{Mp}{\delta}\sqrt{\xi^2-1}\sqrt{1-\eta^2}=\sqrt{\widehat{\Delta}}\sin{\vartheta},\label{BL1}\\
z&=&\frac{Mp}{\delta}\xi\eta=(r-M)\cos{\vartheta},\quad\widehat{\Delta}=r^2-2Mr+M^2 q^2,\label{BL2}
\end{eqnarray}
where strangely enough the deformation parameter completely disappeared from the expressions written in terms of the BL coordinates and the function $\widehat{\Delta}$ is the same as the one entering in the Kerr metric. Moreover, \cite{tomsato} refers to \cite{Voor} for more details but \cite{Voor} provides only the transformations from prolate spheroidal to cylindrical coordinates. Finally, the relations $\rho=\sqrt{\widehat{\Delta}}\sin{\vartheta}$ and $z=(r-M)\cos{\vartheta}$ appear in the work of \cite{Ernst} where Ernst showed how the metric derived therein, is equivalent to the Kerr solution if the coordinate transformation $\xi=(r-M)/Mp$ and $\eta=\cos{\vartheta}$ is introduced. Hence, those parts in the expressions (\ref{BL1}) and (\ref{BL2}) where the $(r,\vartheta)$ coordinates enter, are correct only in the special case $\delta=1$. At this point, we warn the reader about the fact that \cite{Bambi} when analyzing the shadow of a TS manifold with $\delta=2$ made use of the coordinate transformations (\ref{BL1}) and (\ref{BL2}) and therefore, the numerical results presented there should be taken with some caution. The correct transformation for arbitrary $\delta>0$ has been given by \cite{Yam02}, namely
\begin{equation}\label{traff}
r=\frac{Mp}{\delta}\xi+M,\quad \cos{\vartheta}=\eta.
\end{equation}
Note that the condition $\xi\geq 1$ requires that 
\begin{equation}\label{rdelta}
r\geq r_\delta=M+\frac{Mp}{\delta}.
\end{equation}
Then, it is straightforward to verify that the cylindrical coordinates can be expressed as
\begin{equation}\label{minchia}
\rho=\sqrt{(r-M)^2-\left(\frac{Mp}{\delta}\right)^2}\sin{\vartheta},\quad z=(r-M)\cos{\vartheta}.
\end{equation}
Note that $r=r_\delta$ corresponds to $\rho=0$ and $-Mp/\delta\leq z\leq Mp/\delta$, that is to an infinitesimally thin rod of length $2Mp/\delta$. In the present work, we are interested in the case $\delta=2$. Hence,  (\ref{minchia}) becomes
\begin{eqnarray}
\rho&=&\frac{Mp}{2}\sqrt{\xi^2-1}\sqrt{1-\eta^2}=\sqrt{\Delta}\sin{\vartheta},\label{BL3}\\
z&=&\frac{Mp}{2}\xi\eta=(r-M)\cos{\vartheta},\quad
\Delta=(r-M)^2-\frac{M^2 p^2}{4}.\label{BL4}
\end{eqnarray}
 It is not difficult to verify that the prolate spheroidal coordinates can be expressed in terms of the BL coordinates as
\begin{eqnarray}
\xi&=&\frac{\sqrt{R_+}+\sqrt{R_-}}{Mp},\quad \eta=\frac{(r-M)\cos{\vartheta}}{\sqrt{R_+}+\sqrt{R_-}},\label{coorxy}\\
R_\pm&=&\Delta\sin^2{\vartheta}+\left[(r-M)\cos{\vartheta}\pm \frac{Mp}{2}\right]^2=\left(r-M\pm\frac{Mp}{2}\cos{\vartheta}\right)^2\label{Rpm}
\end{eqnarray}
and as a consistency check, it can be easily shown that under the condition $r\geq r_2=M+Mp/2$, formulae (\ref{coorxy}) and (\ref{Rpm}) give
\begin{equation}\label{cap}
\xi=\frac{2}{Mp}(r-M),\quad\eta=\cos{\vartheta}
\end{equation}
in agreement with (\ref{traff}). Finally, the TS line element in BL coordinates is 
\begin{equation}\label{BLTS}
ds^2=\frac{A}{B}dt^2-4Mq\frac{C}{B}\sin^2{\vartheta}dtd\varphi-\Sigma_1\left(\frac{dr^2}{\Delta}+d\vartheta^2\right)-\Sigma_2 \sin^2{\vartheta}d\varphi^2
\end{equation}
with
\begin{equation}
\Sigma_1=\frac{M^8 p^4 B}{256\left[(r-M)^2-\frac{M^2 p^2}{4}\cos^2{\vartheta}\right]^3},\quad
\Sigma_2=\frac{1}{A}\left(B\Delta-4M^2 q^2\frac{C^2}{B}\sin^2{\vartheta}\right)
\end{equation}
and $A$, $B$ and $C$ given by (\ref{Acof}), (\ref{Bcof}) and (\ref{Cof}) where $\xi$ and $\eta$ must be replaced according to (\ref{cap}). To find the radii of the ergoregion on the equatorial plane, we can use (\ref{ergoxi}) with $\eta=0$ and (\ref{coorxy}) with $\vartheta=\pi/2$ together with the constraint $r\geq r_2$ where $r_2$ is defined according to (\ref{rdelta}) to obtain the formulae
\begin{equation}\label{raggio}
r_{\mathcal{E},e}^\pm=M\left(1+ \frac{p}{2}\xi^\pm_{\mathcal{E},e}\right)
\end{equation}
with $\xi^\pm_{\mathcal{E},e}$ given as in (\ref{nunc}). Here, the subscript $e$ stands for equator. For typical numerical values of these radii we refer to Table~{\ref{T2}}. In Fig.~\ref{f02} we represented the ergoregion for the case $q=0.5$ where cusp singularities emerge at the top and bottom of the inner surface of the ergoregion.

At this point, some comments are in order. First of all, $r=r_{\mathcal{E},e}^{-}$ remains a curvature singularity also after we performed the coordinate transformation on the TS metric. Moreover, according to the discussion here below, it turns out that switching to Boyer-Lindquist coordinates has the effect of introducing an event horizon located in the inner region of the ring singularity. This behaviour is not surprising if we keep in mind the example of the Schwarzschild metric. In Kruskal-Szekeres coordinates there is no horizon but if we rewrite the line element in spherical coordinates an horizon appears at $r=2M$. In order to proceed further, we recall that an event horizon can be characterized in terms of a null surface $\mathcal{S}$, that is $g^{\alpha\beta}\partial_\alpha\mathcal{S}\partial_\beta\mathcal{S}=0$ \cite{Islam}. In the case of the TS metric in Boyer-Lindquist coordinates such an equation reduces to
\begin{equation}\label{null}
\left[(r-M)^2-\frac{M^2 p^2}{4}\cos^2{\vartheta}\right]^3\left[\Delta\left(\partial_r\mathcal{S}\right)^2+\left(\partial_\vartheta\mathcal{S}\right)^2\right]=0.
\end{equation}
Let us consider the surface $\Delta=0$, i.e. $r=M\pm Mp/2$. Call these surfaces $\mathcal{S}_+$ and $\mathcal{S}_{-}$. It can be easily verified that these surfaces 
indeed satisfy (\ref{null}) so that they are null surfaces. However, due to the condition $r\geq r_2$ only the surface $\mathcal{S}_+$ should be taken into account. Since $\xi^{-}_{\mathcal{E},e}>1$ when $q>0$ (see Table~\ref{T1}), it follows from (\ref{raggio}) that $r_{\mathcal{E},e}^{-}>r_2$. Hence, in addition to the ring-shaped naked singularity there is an event horizon placed in the inner region inside the aforementioned curvature singularity. However, such a null surface corresponds to the segment $-Mp/2\leq z\leq Mp/2$. Finally, we observe that the term in the first bracket of (\ref{null}) vanishes at
\begin{equation}
r_\pm=M\pm\frac{Mp}{2}\cos{\vartheta}.
\end{equation}
with $r_+$ and $r_{-}$ coinciding with $r_2$ for $\vartheta=0$ and $\vartheta=\pi$, respectively, and otherwise, $r_\pm<r_2$. Since $\Delta$ becomes purely imaginary on the interval $r_{-}<r<r_{+}$, we can neglect this case. 

\section{Conclusions and Outlook}
Even though the TS space-time is relatively unknown to the majority of researchers in General Relativity and its application in astrophysics might be questionable as pointed out by \cite{gary}, we decided to focus the present work on the TS metric due mainly to two reasons. The first one is related to the observation that the literature on this topic is littered with mistakes and misprints which have been able to propagate  from the seminal work of \cite{tomsato,tomsatop} to the most recent result concerning the shadow of a gravitational object described by the TS metric (see \cite{Bambi}). Hence, the first merit of the present work is to have uncovered and corrected several mistakes in the literature. In doing that, we were able to obtain an exact expression for the ergoregion boundary of the TS space-time which in turn allowed to derive exact formulae of the radii of the ergoregion in prolates spheroidal coordinates and in Boyer-Lindquist coordinates. In addition, we also discovered that there is an event horizon located at the inner region of a ring-shaped naked singularity.

The second reason behind our work is related to the claims made in \cite{Bose} where the authors, beside presenting the derivation of a weak lensing formula recently corrected by \cite{BG}, analyse the motion of a massive test particle starting at space-like infinity with zero total angular momentum. It was found by \cite{Bose} that for $q=0$ and $q=1$ the corresponding trajectories emerging from the geodesic equations exhibit a perihelion. However, \cite{Bose} was not able to show whether the same happens for all intermediate values of the parameter $q$. Hence, \cite{Bose} conjectured that the conclusion should also hold for $0<q<1$. In order to prove/disprove such a conjecture in a forthcoming paper, it is desirable to have bulletproof formulae to rely on and therefore, the present work represents a first step in this direction.

Moreover, in view of the fact that the capabilities of very long baseline interferometry have improved in the recent years and are expected to further increase in the next decade \cite{VB1,VB2} at the extent of allowing the observation of the accreation disk around a candidate black hole, a strong gravitational lensing analysis of the TS metric might provide a valuable theoretical resource for the experimental search of such a gravitational object in the observed universe. A first step in this direction would require the (in)-stability analysis of null circular orbits for the TS metric. Should they emerge from a saddle point in the effective potential, then extra care needs to be applied  in order to find out whether or not such orbits are (un)-stable. Preliminary computations show that in principle this problem can be reasonably solved using techniques similar to those adopted in \cite{Cm1,Cm2}.

Last but not least, the TS metric is asymptotically flat while the cosmological data collected in the last decades seem to suggest that our universe has a positive cosmological constant. On the other hand, to the best of the author's knowledge, no study has been conducted so far in the literature to extend the TS metric in the presence of a positive cosmological constant. Therefore,it would be interesting to address the problem of deriving the deSitter-TS metric in a forthcoming paper. Excellent starting points for such an endeavour are represented by \cite{Ch,As}.

\begin{table}[H]
\caption{Typical values of the horizon $r_2$ and the inner and outer boundaries of the ergoregion $r_{\mathcal{E},e}^\pm$ on the equatorial plane.}
\begin{center}
\begin{tabular}{ | l | l | l | l|l|l|l|l|}
\hline
$\eta=0$  &$r_{2}/M$      &$r^{-}_{E,e}/M$     &$r^{+}_{E,e}/M$ \\ \hline
$q=0.1$    &1.497493719   &1.498119109         &1.592179268     \\ \hline
$q=0.2$    &1.489897949   &1.492404180         &1.651650002     \\ \hline
$q=0.3$    &1.476969601   &1.482625988         &1.703951892     \\ \hline
$q=0.4$    &1.458257570   &1.468355754         &1.752205971     \\ \hline
$q=0.5$    &1.433012702   &1.448872522         &1.797677529     \\ \hline
$q=0.6$    &7/5           &1.422966806         &1.841040496     \\ \hline
$q=0.7$    &1.357071421   &1.388483971         &1.882710084     \\ \hline
$q=0.8$    &13/10         &1.341040496         &1.922966806     \\ \hline
$q=0.9$    &1.217944947   &1.268802898         &1.962012653     \\ \hline
$q=0.99$   &1.070533680   &1.117858921         &1.996245294     \\ \hline
\end{tabular}
\label{T2}
\end{center}
\end{table}

\appendix
\section{Numerical tables}\label{numeric}
We present here the numerical values for the inner and outer boundaries $\xi_\mathcal{E}^\pm$ of the ergoregion for different values of the rotational parameter $q$ and of the angular variable $\eta$ (see Tables~\ref{App1},~\ref{App4} and \ref{App7}). More precisely, we considered the intervals $0.1\leq q\leq 0.99$ and $0.1\leq\eta\leq 0.99$. In these ranges, the roots of the polynomial function $B$ entering the metric coefficient $f$ are always complex. This indicates that modulo the quasi-regular singularities at $(1,\pm 1)$, $f$ can only become singular on the equatorial plane.
\begin{table}[H]
\caption{Typical values of $\xi_\mathcal{E}^\pm$ in prolate spheroidal coordinates for $0.1\leq q\leq 0.99$ and $0.1\leq\eta\leq 0.3$. }
\begin{center}
\begin{tabular}{ | l | l | l | l|l|l|l|}
\hline
$\eta=0.1$ &$\xi_\mathcal{E}^{-}$ & $\xi_\mathcal{E}^+$     \\ \hline
$q=0.1$    &1.001244519 &1.188572734    \\ \hline
$q=0.2$    &1.005064795 &1.327279875    \\ \hline
$q=0.3$    &1.011741110 &1.471887154    \\ \hline
$q=0.4$    &1.021818037 &1.636278712    \\ \hline
$q=0.5$    &1.036266826 &1.835649270    \\ \hline
$q=0.6$    &1.056858285 &2.094450443    \\ \hline
$q=0.7$    &1.087128247 &2.461723165    \\ \hline
$q=0.8$    &1.135515223 &3.062767535    \\ \hline
$q=0.9$    &1.231237701 &4.393029551    \\ \hline
$q=0.99$   &1.665588339 &14.05394768    \\ \hline
\end{tabular}
\label{App1}
\quad
\begin{tabular}{ | l | l | l | l|l|l|l|}
\hline
$\eta=0.2$ &$\xi_\mathcal{E}^{-}$ & $\xi_\mathcal{E}^+$     \\ \hline
$q=0.1$    &1.001206829 &1.183300016    \\ \hline
$q=0.2$    &1.004911691 &1.318556370    \\ \hline
$q=0.3$    &1.011387321 &1.459830391    \\ \hline
$q=0.4$    &1.021163733 &1.620671921    \\ \hline
$q=0.5$    &1.035186500 &1.815985114    \\ \hline
$q=0.6$    &1.055180322 &2.069805474    \\ \hline
$q=0.7$    &1.084590838 &2.430379632    \\ \hline
$q=0.8$    &1.131647154 &3.021024505    \\ \hline
$q=0.9$    &1.224872051 &4.329457508    \\ \hline
$q=0.99$   &1.649369985 &13.84046576    \\ \hline
\end{tabular}
\label{App2}
\quad
\begin{tabular}{ | l | l | l | l|l|l|l|}
\hline
$\eta=0.3$ &$\xi_\mathcal{E}^{-}$ & $\xi_\mathcal{E}^+$     \\ \hline
$q=0.1$    &1.001144009 &1.174459550    \\ \hline
$q=0.2$    &1.004656467 &1.303887504    \\ \hline
$q=0.3$    &1.010797398 &1.439511398    \\ \hline
$q=0.4$    &1.020072295 &1.594321140    \\ \hline
$q=0.5$    &1.033383445 &1.782729578    \\ \hline
$q=0.6$    &1.052377772 &2.028065112    \\ \hline
$q=0.7$    &1.080348579 &2.377222226    \\ \hline
$q=0.8$    &1.125170823 &2.950140542    \\ \hline
$q=0.9$    &1.214188473 &4.221377158    \\ \hline
$q=0.99$   &1.621979060 &13.47714996    \\ \hline
\end{tabular}
\label{App3}
\end{center}
\end{table}
\begin{table}[H]
\caption{Typical values of $\xi_\mathcal{E}^\pm$ in prolate spheroidal coordinates for $0.1\leq q\leq 0.99$ and $0.4\leq\eta\leq 0.6$. }
\begin{center}
\begin{tabular}{ | l | l | l | l|l|l|l|}
\hline
$\eta=0.4$ &$\xi_\mathcal{E}^{-}$ & $\xi_\mathcal{E}^+$     \\ \hline
$q=0.1$    &1.001056055 &1.161969906    \\ \hline
$q=0.2$    &1.004299043 &1.283069381    \\ \hline
$q=0.3$    &1.009970927 &1.410573173    \\ \hline
$q=0.4$    &1.018542316 &1.556680857    \\ \hline
$q=0.5$    &1.030853871 &1.735101349    \\ \hline
$q=0.6$    &1.048441617 &1.968141982    \\ \hline
$q=0.7$    &1.074381275 &2.300739449    \\ \hline
$q=0.8$    &1.116040824 &2.847940559    \\ \hline
$q=0.9$    &1.199071557 &4.065240094    \\ \hline
$q=0.99$   &1.582835645 &12.95139687    \\ \hline
\end{tabular}
\label{App4}
\quad
\begin{tabular}{ | l | l | l | l|l|l|l|}
\hline
$\eta=0.5$ &$\xi_\mathcal{E}^{-}$ & $\xi_\mathcal{E}^+$     \\ \hline
$q=0.1$    &1.000942959 &1.145711749    \\ \hline
$q=0.2$    &1.003839312 &1.255796218    \\ \hline
$q=0.3$    &1.008907327 &1.372470620    \\ \hline
$q=0.4$    &1.016571816 &1.506905323    \\ \hline
$q=0.5$    &1.027592411 &1.671872531    \\ \hline
$q=0.6$    &1.043359026 &1.888305850    \\ \hline
$q=0.7$    &1.066659976 &2.198498272    \\ \hline
$q=0.8$    &1.104191334 &2.710885041    \\ \hline
$q=0.9$    &1.179350835 &3.855211805    \\ \hline
$q=0.99$   &1.531038595 &12.24229644    \\ \hline
\end{tabular}
\label{App5}
\quad
\begin{tabular}{ | l | l | l | l|l|l|l|}
\hline
$\eta=0.6$ &$\xi_\mathcal{E}^{-}$ & $\xi_\mathcal{E}^+$     \\ \hline
$q=0.1$    &1.000804714 &1.125521784    \\ \hline
$q=0.2$    &1.003277132 &1.221635761    \\ \hline
$q=0.3$    &1.007605846 &1.324413020    \\ \hline
$q=0.4$    &1.014158229 &1.443739929    \\ \hline
$q=0.5$    &1.023592070 &1.591184450    \\ \hline
$q=0.6$    &1.037113145 &1.785888145    \\ \hline
$q=0.7$    &1.057146248 &2.066679326    \\ \hline
$q=0.8$    &1.089533626 &2.533323122    \\ \hline
$q=0.9$    &1.154790469 &3.581824704    \\ \hline
$q=0.99$   &1.465246589 &11.31543172    \\ \hline
\end{tabular}
\label{App6}
\end{center}
\end{table}
\begin{table}[H]
\caption{Typical values of $\xi_\mathcal{E}^\pm$ in prolate spheroidal coordinates for $0.1\leq q\leq 0.99$ and $0.7\leq\eta\leq 0.9$. }
\begin{center}
\begin{tabular}{ | l | l | l | l|l|l|l|}
\hline
$\eta=0.7$ &$\xi_\mathcal{E}^{-}$ & $\xi_\mathcal{E}^+$     \\ \hline
$q=0.1$    &1.000641309 &1.101183763    \\ \hline
$q=0.2$    &1.002612332 &1.179990007    \\ \hline
$q=0.3$    &1.006065563 &1.265266537    \\ \hline
$q=0.4$    &1.011298381 &1.365327537    \\ \hline
$q=0.5$    &1.018844145 &1.490203770    \\ \hline
$q=0.6$    &1.029682805 &1.656706163    \\ \hline
$q=0.7$    &1.045791176 &1.899133583    \\ \hline
$q=0.8$    &1.071952509 &2.305916399    \\ \hline
$q=0.9$    &1.125073556 &3.229025779    \\ \hline
$q=0.99$   &1.383464435 &10.11109287    \\ \hline
\end{tabular}
\label{App7}
\quad
\begin{tabular}{ | l | l | l | l|l|l|l|}
\hline
$\eta=0.8$ &$\xi_\mathcal{E}^{-}$ & $\xi_\mathcal{E}^+$     \\ \hline
$q=0.1$    &1.000452731 &1.072415310    \\ \hline
$q=0.2$    &1.001844706 &1.130031675    \\ \hline
$q=0.3$    &1.004285376 &1.193383547    \\ \hline
$q=0.4$    &1.007988474 &1.268844574    \\ \hline
$q=0.5$    &1.013338126 &1.364432382    \\ \hline
$q=0.6$    &1.021042148 &1.493832491    \\ \hline
$q=0.7$    &1.032534010 &1.685238687    \\ \hline
$q=0.8$    &1.051301327 &2.011832721    \\ \hline
$q=0.9$    &1.089778339 &2.766600056    \\ \hline
$q=0.99$   &1.282637129 &8.512310772    \\ \hline
\end{tabular}
\label{App8}
\quad
\begin{tabular}{ | l | l | l | l|l|l|l|}
\hline
$\eta=0.9$ &$\xi_\mathcal{E}^{-}$ & $\xi_\mathcal{E}^+$     \\ \hline
$q=0.1$    &1.000238967 &1.038848419    \\ \hline
$q=0.2$    &1.000974018 &1.070597706    \\ \hline
$q=0.3$    &1.002264010 &1.106284089    \\ \hline
$q=0.4$    &1.004224058 &1.149750752    \\ \hline
$q=0.5$    &1.007061581 &1.206139834    \\ \hline
$q=0.6$    &1.011160146 &1.284514335    \\ \hline
$q=0.7$    &1.017300392 &1.403968178    \\ \hline
$q=0.8$    &1.027394968 &1.615050284    \\ \hline
$q=0.9$    &1.048341544 &2.124117081    \\ \hline
$q=0.99$   &1.157799663 &6.222113920    \\ \hline
\end{tabular}
\label{App9}
\end{center}
\end{table}


\begin{thebibliography}{99}
\bibitem{tomsato} 
A. Tomimatsu, and H. Sato, {\it New exact solution for the gravitational field of a spinning mass}, Phys. Rev. Lett. {\bf 29}, 1344 (1972).
\bibitem{tomsatop} 
A. Tomimatsu, and H. Sato, {\it New series of exact solutions for gravitational fields of spinning masses}, Progr. Theor. Phys. {\bf 50}, 95  (1973).
\bibitem{Yam01} 
M. Yamazaki, {\em On the Kerr and the Tomimatsu-Sato Spinning Mass Solutions}, Progr. Theor. Phys. {\bf{57}}, 1951 (1977).
\bibitem{Yam02} 
M. Yamazaki, {\em On the Kerr-Tomimatsu-Sato family of spinning mass solutions}, J. Math. Phys. {\bf{18}}, 2502 (1977).
\bibitem{Hori}
S. Hori, {\it{On the Exact Solution of Tomimatsu-Sato Family for an Arbitrary Integral Valued of the Deformation Parameter}}, Progr. Theor. Phys. {\bf{59}}, 1870 (1978); Erratum-ibid. {\bf{61}}, 365 (1979).
\bibitem{Tani}
O. Tanimura, {\it{Solution of the Ernst Equation for a Real Value of the Deformation Parameter}}, Progr. Theor. Phys. {\bf{100}}, 523 (1998).
\bibitem{gary}
G. W. Gibbons and R. A. Russell-Clark, {\it{Note on the Tomimatsu-Sato Solution of Einstein's Equations}}, Phys. Rev. Lett. {\bf{30}}, 398 (1973).
\bibitem{CH}
S. Chandrasekhar, {\it{The Mathematical Theory of Black Holes}}, (Oxford University Press, 1983).
\bibitem{Penrose}
R. Penrose, {\it{Gravitational collapse: The role of general relativity}}, Riv.Nuovo Cim. {\bf{1}} 252 (1969),  Gen.Rel.Grav. {\bf{34}}, 1141-1165 (2002).
\bibitem{Papa1}
A. Papapetrou, {\it{A Randowm Walk in General Relativity}}, Wiley \& Sons, New York (1985).
\bibitem{Hollier}
G. P. Hollier, {\it{Papapetrou's naked singularity is a strong curvature singularity}}, Class. Quantum Grav. {\bf{3}}, L111 (1986)
\bibitem{Singh}
T. P. Singh, {\it{Gravitational collapse and cosmic censorship}}, IMSC Report {\bf{117}}, 57 (1996).
\bibitem{Gundlach}
C. Gundlach, {\it{Critical phenomena in gravitational collapse}}, Living Rev. Rel. {\bf{10}}, 5 (2007).
\bibitem{Harada}
T. Harada, H. Iguchi and K.I. Nakao, {\it{Physical Processes in Naked Singularity Formation}}, Prog. Theor. Phys. {\bf{107}}, 449 (2002).
\bibitem{Boyda}
E. K. Boyda, S. Ganguli, P. Ho\v{r}ava and U. Varadarajan, {\it{Holographic protection of chronology in universes of the G\"{o}del type}}, Phys. Rev. D {\bf{67}}, 106003 (2003).
\bibitem{Israel}
D. Israel, {\it{Quantization of heterotic strings in a Godel/anti-de Sitter space-time and chronology protection}}, JHEP {\bf{01}}, 042 (2004).
\bibitem{Drukker}
N. Drukker, {\it{Supertube domain-walls and elimination of closed time-like curves in string theory}}, Phys. Rev. D {\bf{70}}, 084031 (2004).
\bibitem{Lewis}
T. Lewis, {\it{Some Special Solutions of the Equations of Axially Symmetric Gravitational Fields}}, Proc. R. Soc. London A {\bf{136}}, 176 (1932).
\bibitem{Papa}
A. Papapetrou, {\it{Eine rotationssymmetrische L\"{o}sung in der allgemeinen Relativit\"{a}tstheorie}}, Annals Phys. {\bf{12}}, 309 (1953).
\bibitem{Islam}
J. N. Islam, {\it{Rotating fields in general relativity}}, Cambridge University Press, Cambridge (2009).
\bibitem{Cosgrove1}
C. M. Cosgrove, {\it{A new formulation of the field equations for the stationary axisymmetric vacuum gravitational field. I. General theory}}, J. Phys. A: Math. Gen. {\bf{11}}, 2389 (1978).
\bibitem{Cosgrove2}
C. M. Cosgrove, {\it{A new formulation of the field equations for the stationary axisymmetric vacuum gravitational field. II. Separable solutions}}, J. Phys. A: Math. Gen. {\bf{11}}, 2405 (1978).
\bibitem{Ernst}
F. J. Ernst, {\em New Formulation of the Axially Symmetric Gravitational Field Problem}, Phys. Rev. {\bf{167}}, 1175 (1968).
\bibitem{Voor}
B. H. Voorhees, {\it{Static Axially Symmetric Gravitational Fields}}, Phys. Rev. D {\bf{2}}, 2119 (1970).
\bibitem{Moon}
P. Moon and D. E. Spencer, {\it{Field Theory Handbook}}, 2d ed., Berlin: Springer-Verlag (1971)
\bibitem{Kodama}
H. Kodama and W. Hikida, {\it{Global structure of the Zipoy-Voorhees-Weyl spacetime and the $\delta=2$ Tomimatsu-Sato spacetime}}, Class. Quantum Grav. {\bf{20}}, 5121 (2003).
\bibitem{Wald}
R. M. Wald, {\it{General Relativity}}, University of Chicago Press (1984).
\bibitem{Bron}
I. N. Bronstein, K. A. Semendjajew, G. Musiol and H. M\"{u}lig {\it{Taschenbuch der Mathematik}}, Verlag Harri Deutsch, Frankfurt am Main (2005).
\bibitem{Bambi}
C. Bambi and N. Yoshida, {\it{Shape and position of the shadow in the $\delta=2$ Tomimatsu-Sato space-time}}, Class. Quant. Grav. {\bf{27}}, 205006 (2010).
\bibitem{papa90}
D. Papadopoulos and B. C. Xanthopoulos, {\it{Tomimatus-Sato solutions describe cosmic strings interacting with gravitational waves}}, Phys. Rev. D {\bf{41}}, 2512 (1990).
\bibitem{Bose}
S.K. Bose and M. Y. Wang, {\it{Geodesic Motions in the Tomimatsu-Sato Metric}}, Phys. Rev. D {\bf{8}}, 361 (1973).
\bibitem{BG}
D. Batic, S.Chanda and P. Guha, {\it{Optical properties of null geodesics emerging from dynamical systems}}, Eur. Phys. J. C {\bf{82}}, 422 (2022).
\bibitem{VB1}
 S. Doeleman et al., {\it{Event-horizon-scale structure in the supermassive black hole candidate at the Galactic Centre}}, Nature {\bf{455}}, 78 (2008).
\bibitem{VB2}
S. Doeleman et al., {\it{Imaging an Event Horizon: submm-VLBI of a Super Massive Black Hole}}, Science White Paper submitted to the ASTRO2010 Decadal Review Panels,  
 arXiv:0906.3899 (2009).
\bibitem{Cm1}
M. Alrais Alawadi, D. Batic and M. Nowakowski, {\it{Light bending in a two black hole metric}}, Class. Quantum Grav. {\bf{38}}, 045003 (2021).
\bibitem{Cm2}
D. Batic, H. A. Kittaneh and M. Nowakowski, {\it{Relevant scales for the $C$-metric with positive cosmological constant}}, Phys. Rev. D {\bf{104}}, 124029 (2021).
\bibitem{Ch}
C. Charmousis, D. Langlois, D. Steer and R. Zegers, {\it{Rotating spacetimes with a cosmological constant}}, JHEP {\bf{0702}}, 064 (2007).
\bibitem{As}
M. Astorino, {\it{Charging axisymmetric space-times with cosmological constant
}}, JHEP {\bf{06}}, 086 (2012).
\end{thebibliography}
\end{document}